\documentclass[times, review, 10pt]{elsarticle}

\usepackage[english]{babel}
\usepackage[utf8x]{inputenc}
\usepackage[T1]{fontenc}

\usepackage{newtxtext}
\usepackage{amsmath,amsfonts}
\usepackage{csquotes}
\usepackage{algorithmic}
\usepackage{array}
\usepackage[caption=false,font=normalsize,labelfont=sf,textfont=sf]{subfig}
\usepackage{textcomp}
\usepackage{stfloats}
\usepackage{url}
\usepackage{verbatim}
\usepackage{graphicx}
\usepackage{color}              
\usepackage{hyperref}
\usepackage{algorithmic}
\usepackage{multirow}
\usepackage[linesnumbered,ruled,vlined]{algorithm2e}
\usepackage{setspace}

\def\tsc#1{\csdef{#1}{\textsc{\lowercase{#1}}\xspace}}
\tsc{WGM}
\tsc{QE}
\DeclareMathOperator*{\argmax}{arg\,max}

\def\BibTeX{{\rm B\kern-.05em{\sc i\kern-.025em b}\kern-.08em
		T\kern-.1667em\lower.7ex\hbox{E}\kern-.125emX}}
\usepackage{balance}

\def\vx{\boldsymbol{x}}
\def\vf{\boldsymbol{f}}
\def\mW{\boldsymbol{W}}
\def\mF{\boldsymbol{F}}
\def\mY{\boldsymbol{Y}}
\def\mS{\boldsymbol{S}}
\def\mD{\boldsymbol{D}}
\def\mI{\boldsymbol{I}}

\begin{document}

\begin{frontmatter}

\title{Graph-Based Semi-Supervised Hyperspectral Image Classification with Distance-Aware Spatial Measure}
    
    \author{S\'ergio~J.~M.~Almeida\textsuperscript{a}}
    \ead{sergio.almeida@ucpel.edu.br}
    
    \author{Jos\'e C. M. Bermudez\textsuperscript{b,*,1}}
    \ead{jose.bermudez@ufsc.br}

    \address{\textsuperscript{a} Catholic University of Pelotas, Center for Social and Technological Sciences, Pelotas, RS, Brazil}
    \address{\textsuperscript{b} Federal University of Santa Catarina, Department of Electrical and Electronic Engineering, Florian\'opolis, SC, Brazil}
    


	
\singlespacing
\begin{abstract}
The classification of hyperspectral images (HIs) still presents several challenges. One of them is the difficulty to obtain a large set of labeled samples to train the classifier. Semi-supervised learning methods have received much attention recently, as they require the initial labeling of a reduced number of image pixels and lead to very good results for practical application. One of the open problems in graph-based semi-supervised HI classification is the need to consider relative spatial relationship between pixels in the image to improve the smoothness of the solution.  Kernel-based approaches using composite kernels have led to very good results, in which one kernel addresses the spectral properties of the pixels while a second kernel addresses some spatial properties. Most available solutions employ a spectral-spatial kernel which considers the spectral properties of a spatial region about each pixel. This work proposes a composite kernel approach that includes a third kernel dealing exclusively with the relative spatial position of the pixels. Experiments with real HI images show that the use of the new spatial kernel has led to improved classification results when compared to those previously reported in the literature. These results also shed some light on the relative contributions of the spectral and spectral-spatial kernels in semi-supervised HI classification. 
\end{abstract}


\end{frontmatter}
\maketitle

\section{Introduction}
Hyperspectral image (HI) processing has been widely employed in several areas, such as land-cover classification, mineral identification, precision agriculture,  medical image analysis, biometric identification, and drug discovery among others. 
Nevertheless, classification tasks employing HIs still present significant challenges \cite{Luo-2019}, \cite{Zhang-2012}, \cite{Tarabalka-2010}.  The classification task becomes even more difficult when small groups of labeled pixels are available. This information-poor scenario makes capturing the underlying probability distribution function of the image a complex task.

HIs contain a large amount of information, which leads to greater accuracy. In this sense, many machine learning methods have been explored for HI classification, such as, unsupervised clustering \cite{Hung-2011}, and supervised classification \cite{Villa-2011}, \cite{Lixia-2014}. However, this large amount of information is also accompanied by a large number of spectral channels, which results in a higher computational cost to obtain the correct labels.  The availability of few samples for training, associated with a high number of spectral channels, can result in the much feared curse of dimensionality, leading to overfitting the model to the training data. 

One factor that complicates HI classification is the intrinsic spectral variability of the materials in the scene. The spectrum of a given material (and thus its HI pixel representation) can be significantly affected by variations in atmospheric, illumination, and environmental conditions that typically occur within an image~\cite{Borsoi-2020a}.  Due to spectral variability, even pixels located in small regions containing the same physical material (hence belonging to the same class) may have reasonably distinct spectral representations.  Another important factor that may arise is a spatial variability of the spectral signatures  \cite{Camps-Valls-2007,  Kun-2015, Gomez-2008, Lixia-2014}. Depending on the configuration of an area being classified, pixels of the same class may be located in distant regions of the image. Such situation is not uncommon and represents an additional challenge for the classification algorithms.

The main difficulty with all supervised methods in the learning process is the need for a good quality training data set, as labeled samples are usually not available in sufficient quantity due to the high cost of labeling. Unsupervised methods may yield satisfactory results in some classification problems because they are not sensitive to the number of labeled samples. However, the relationship between clusters and classes is not assured \cite{Camps-Valls-2007}.

In a realistic scenario, where the data volume is increasingly larger and, at the same time, the available labeled data is scarce, mainly due to labelling costs, the development of semi-supervised learning (SSL) methods has become a natural trend in the search for better classifier performances. SSL is halfway between unsupervised and fully supervised learning. In SSL, the algorithm is provided with the labeled data information (data and labels), the unlabeled data, and some supervision information~\cite[Ch.~1]{Chapelle-2006}. SSL algorithms allow the use of unlabeled data to complement the information provided by the small amount of labeled data to improve classifier performance.

SSL methods consider a data set of $n$ samples $\vx_i$ of an input space $\mathcal{X}$, composed of a labeled data set $\mathcal{D}_l = \{\vx_i, y_i\}|_{i=1,\ldots,n_l}$, where $\vx_i$ are the labeled samples, and $y_i$ are their corresponding labels in $\mathcal{Y}$, and an unlabeled data set $\mathcal{D}_u = \{\vx_i\}|_{i=n_l+1,\ldots,n}$ with $n_u$ samples for which the labels are unknown, such that $n = n_l + n_u$. Mathematically, SSL is useful when the knowledge gained on the probability density function $p(\vx)$ by using the unlabeled data information can be of help in the inference of the conditional density $p(y|\vx)$.

SSL algorithms may be based on generative or discriminative models. The generative approach estimates the posterior class distribution by modeling the conditional class distributions explicitly \cite{Dempster-2018}. The discriminative approach estimates the posterior class distribution directly, without the need to specify the conditional class distributions explicitly \cite{Gomez-2008}. 

At a high level, SSL algorithms can also be classified as employing inductive or transductive methods \cite{LIU-2009, Jesper-2020,Chapelle-2006,Zhu-2008}. Inductive methods attempt to find a classification model that can be used later on to classify new data. Transductive methods obtain label prediction for the given unlabeled data without building a classification model. Within the latter methods we can highlight two subgroups of SSL algorithms: 1) transductive SVM (TSVM) \cite{Bruzzone-2006, Bo-2014, Singla-2021, Geng-2019}, and graph-based methods \cite[Part III]{Chapelle-2006}\cite{chung-1997,Jordan-1999,Song-graph-2023, Tanni-2025, Freire-2025, Culp-2008}. 

Graph-based SSL transductive methods \cite{Jesper-2020}  have gained special attention due to their ability to extract characteristics existing in the graph nodes and to establish relationships between these characteristics in order to facilitate the classification task. This ability has led to their application in several fields \cite{Chapelle-2006}. Each node in the graph corresponds to data, and the edges are defined by weights that measure the similarity between the data in each node pair. This ability of graphs to explore the correlations between nodes has led to the development of semi-supervised learning techniques that address the challenge of having a small set of labeled data. Due to its relevance for the classification performance, the graph construction has a key role in graph-based transductive SSL methods.

Several graph-based SSL algorithms have been proposed for HI classification. For instance, in \cite{Gu-L1-graph-2012}, a sparse set of graph weights are obtained by solving an L1 optimization problem. Then, the SSL classification method proposed in~\cite{Zhou-2004} is applied. The method proposed in~\cite{Shao-discriminant-2017} combines both L1-graph and partial labeled information to construct a discriminant sparse representation (DSR) graph. In such a graph structure, each pairwise nodes are treated differently according to the probability that they belong to the same class. The class-probability of unlabeled samples is estimated using sparse representation based classification, in which each unlabeled sample is encoded as a sparse combination of labeled samples, and the obtained coefficients are interpreted as the similarity between the unlabeled sample and each labeled sample. In this work we are interested in those methods that address both the spectral and spatial variability problems. 

In supervised methods, several works have proposed different strategies to combine spectral and spatial information to improve HI classification. One of the most popular approaches is to obtain an initial coarse classification based only on the spectral information, and then refine the pixel labeling using conditional random fields (CRF) defined over pixels or image regions.

The work in \cite{krahenbuhl2011efficient} addresses the design of CRFs defined over pixels or image regions for the multi-class supervised image segmentation and labeling. The paper proposes an approximate inference algorithm for fully connected CRF models that establishes pairwise potentials on all pairs of pixels in the image using a linear combination of Gaussian kernels. The proposed fully connected pairwise CRF model is characterized by a Gibbs distribution with Gibbs energy composed by a unary potential computed independently for each pixel by a classifier that produces a distribution over the label assignment given image features, and by pairwise potentials defined by a weighted sum of Gaussian kernels. For multi-class image segmentation and labeling the authors propose contrast-sensitive two-kernel potentials defined in terms of the color vectors and positions of each pair of pixels. These composite kernels include contributions determined by the Gaussian kernel applied to the difference of the pixel positions. 

In hyperspectral supervised image classification, CRF is traditionally used to perform image segmentation after an initial coarse pixel-level class label has been generated. The CRF is designed so that pixels in a local neighborhood tend to have the same class label. Then, the outcome of CRF can be considered as an improved classification map. Several recent methods integrate convolutional neural networks (CNNs) and CRF methods \cite{liang2015semantic, pan2018high, alam2016crf, liang2019hyperspectral}.  In these methods, the pixel positions are used in the definition of pairwise potentials as a weighted sum of the appearance and smooth Gaussian kernels associated to each pair of pixels. The pairwise potentials are then combined with the unary potentials (resulting from the initial CNN classifications) to build the energy function of the CRF. More recently, \cite{alam2018conditional} proposed to further employ a specific deconvolutional network to produce the final classifications. All these works consider supervised classification. Also, the implementation of these methods can become quite complex and require a significant computational effort.

Most SSL methods that address spectral-spatial classification do it by employing composite kernels to measure the similarity of graph nodes as an efficient way to circumvent the curse of dimensionality~\cite{Camps-Valls-2007}. 

The work \cite{Camps-Valls-composite-2006} proposed a framework of composite kernel machines for enhanced classification of hyperspectral images. The proposed method exploits the properties of Mercer’s kernels to construct a family of composite kernels that combine spatial and spectral information. The composite kernels are built as a weighted sum of kernels specialized on spatial or spectral features. Every pixel $\vx_i$ in the HI is redefined in the spectral and in the spatial domain. The spectral content is represented by a vector $\vx_i^{\omega}$ $\in \mathbb{R}^{N_{\omega}}$, which is usually  the HI pixel itself. The spatial content is obtained through some type of transformation of the HI pixels in the spatial neighborhood of $\vx_i$, which yields vectors $\vx_i^s$ $\in \mathbb{R}^{N_s}$. In \cite{Camps-Valls-composite-2006} it was proposed that $\vx_i^s$ be evaluated as the average of the reflectance values in a given window around pixel $\vx_i$ for each band. Then, the similarity between two graph signals (nodes) $\vx_i$ and $\vx_j$ is determined by a composite kernel that combines the contributions of the two representations of each pixel. 

The work in~\cite{Camps-Valls-2007} proposed a method based on the transductive SSL algorithm introduced in \cite{Zhou-2004}, but applied in the context of HI. The employed measures of similarity between the graph nodes were kernel-based to alleviate the problems inherent to the high dimensionality of HI data, and used the composite kernels proposed in~\cite{Camps-Valls-composite-2006}. Different composite kernels were employed to account for spectral and spatial variability.	

More recent works have suggested alternative methods for determining the spatial information relative to pixel $\vx_i$, most of them building upon the same idea proposed in \cite{Camps-Valls-2007}. In \cite{Velasco-improving-2009},  after preprocessing through nonlinear diffusion partial differential equations (PDEs) and through wavelet shrinkage, the spatial components $\vx_i^s$ were selected from the smoothed image. The weighted summation kernel originally proposed in \cite{Camps-Valls-composite-2006} was then employed. The graph weighted adjacent matrix, however, would have nonzero elements only if the pixel $j$ belonged to the $k$-nearest neighbors of the pixel $i$. The classification accuracy results reported in \cite{Velasco-improving-2009} are comparable to those obtained using the simpler method of \cite{Camps-Valls-composite-2006}, despite the extra complexity required by the preprocessing step.

In~\cite{Jamshidpour-graph-based-2016}, a graph-based method was proposed in which two graphs are constructed to independently represent the spatial and spectral information. The spectral graph is built by connecting each pixel to its $K$ nearest neighbors in spectral space. The spatial-based graph is built by computing the spectral distance between each pixel and a set of its spatial neighbors. The Laplacians of the two graphs are then combined in a weighted summation and the classification problem is solved as in~\cite{Zhou-2004}.

In~\cite{Ma-graph-based-2016}, a sum of minimum distance (SMD) was proposed to measure the distance between two sets of points defined in a square window around each HI pixel. These distances are then used to build a $kNN$ sparse graph. Next, local manifold learning (LML) is employed as a measure of similarity between two finite sets of points to determine the weights associated with the graph edges. The classification is implemented using the Gaussian ﬁelds and harmonic functions (GFHF) method~\cite{Zhu-semi-supervised-2003}.

Most existing graph-based SSL approaches for HI classification using spectral-spatial information employ basically the composite kernel-based similarity measure proposed in \cite{Camps-Valls-composite-2006}. The spatial representation $\vx_i^s \in \mathbb{R}^{N_s}$, is generated through a feature extraction method that combines contributions of spectral representations of pixels in the area surrounding $\vx_i$ in the HI. Hence, the spatial information embedded in $\vx_i^s$ refers to the surroundings of $\vx_i$ in the spectral feature space. This approach has the property that $\vx_i^s$ will differ from $\vx_i$ if the pixels close to it in the HI have significantly different spectra, compared to $\vx_i$. We note that this approach does not address one important aspect in classification of HIs, in which pixels belonging to a given class may be located in spatially distinct regions of the HI. This is a very common situation in, for instance, classifications of land-covers in remote-sensing applications. In such application, the physical distance between two pixels of which one wants to evaluate the similarity can be leveraged to compensate, for instance, for a lack of similarity of two spatially closed pixels of the same class whose representations differ due to spectral variability. It can also help when pixels from different classes have spectra relatively close in the spectral feature space but are located at very distinct regions of the HI. The work in \cite{Lanckriet-statistical-2004} considered for the first time a multiple kernel learning framework that combined different types of data for integrating heterogeneous descriptions of the same set of genes. Inspired by \cite{Lanckriet-statistical-2004}, by the successful use of the pixel positions in supervised learning \cite{liang2015semantic, pan2018high, alam2016crf, liang2019hyperspectral} and by the considerations above about the importance to include the physical distance between pixels in measuring their similarity, we propose the use of a new spatial term in composite kernels for classification of HIs. Our consideration of the spatial information is based on the coordinates of the pixels within the image. However, differently from the supervised methods described above, we insert the spatial information directly in the definition of a composite spectral-spatial kernel, leading to an SSL algorithm that is simple to apply and leads to very good classification results.  

This paper is organized as follows. Section~\ref{sec_Method} describes the proposed composite kernel structure, which combines the spectral and spatial-spectral kernels with the newly proposed spatial-2D kernel. A relationships among the three kernel bandwidths is also proposed. Section~\ref{sec_Solution} describes the solution of the classification problem using the composite kernel. A step-by-step description of the proposed algorithm is described.  Section~\ref{sec_Simulations} describes five experiments which compare the results obtained using the proposed method with results reported in the literature. The conclusions of the work are presented in Section~\ref{sec_Conclusions}.

\section{Graph-Based Method Proposed} \label{sec_Method}

\subsection{Construction of the spatial-spectral kernel}

Composite kernels were proposed in~\cite{Camps-Valls-composite-2006} and employed in~\cite{Camps-Valls-2007} to consider a spatial contribution in the definition of similarity of two vertices of the graph in graph-based HI classification. These composite kernels included, in fact, two contributions determined in the spectral domain of the pixels. The spectral component was the pixel representation itself, which is a vector whose components are the contributions of the different frequency bands to the spectral representation of the HI pixel. The spatial component of a pixel $\vx_i$ was constructed by averaging the spectral representations of the pixels in a window surrounding $\vx_i$. In these papers it was proposed to take the average of these pixels as the spatial contribution $\vx_i^s$. More recent works proposed different (and more complex) transformations of the pixels surrounding $\vx_i$ to obtain $\vx_i^s$. Such transformations, however, were also built in the spectral domain. 

Inspired by the work in~\cite{Lanckriet-statistical-2004} and by the successful use of pixel positions in Gaussian kernels in the realm of CRF-based supervised HI classification \cite{liang2015semantic, pan2018high, alam2016crf, liang2019hyperspectral}, we propose to construct the kernel-based similarity measure using heterogeneous data. Without disregarding the previous contributions, we propose to add a new term to the composite kernel which affects the similarity of two HI pixels by considering the actual physical distance between these pixels. To this end, we employ an additional spatial pixel representation termed $\vx_i^d \in \mathbb{R}^2$ whose components are the spatial coordinates of the pixel $\vx_i$ in the HI.    

In this paper we concentrate on a weighted-summation composite kernel for simplicity. Extension to other types of compositions should be straightforward.

Consider an HI with an $N$-dimensional input spectral space, $\mathcal{X}=\{\vx_1, \cdots, \vx_{n_l}, \vx_{n_l+1}, \allowbreak \cdots,\vx_n\}$ $\subset$ $\mathbb{R}^N $. The first $n_l$ pixels $\vx_i$ ($i \leq n_l$) are labeled with corresponding labels $y_i$, $i=1,\ldots,n_l$, and the $n_{u}$ pixels $\vx_{i}$ ($n_{l+1} \leq i \leq n$) are unlabeled. 

The proposed composite kernel is built from three distinct pixel representations of a HI pixel $\vx_i$:

\begin{itemize}
	\item[a)] \emph{Pixel definitions}
	
	As in~\cite{Camps-Valls-2007}, the spectral pixel entity $\vx_i^{\omega} \in \mathbb{R}^{N_{\omega}}$ is defined as the HI pixel itself, whose components are the contributions of each frequency band to the pixel value. Thus, $N_{\omega} = N$. The spatial-spectral pixel entity $\vx_i^s \in \mathbb{R}^{N_s}$ is defined using a feature extraction method to the area surrounding $\vx_i$.  In this work we consider the average of the pixels in a window of size $w \times w$ around $\vx_i$, which corresponds to evaluate the average per spectral band~\cite{Camps-Valls-composite-2006}. Hence, $N_s = N_{\omega} = N$. Finally, we define a new spatial-2D pixel entity $\vx_i^d \in \mathbb{R}^2$ given by the coordinate vector of $\vx_i$ in the HI.\\
	
	\item[b)] \emph{Kernel computation}
	
	Given the three entities defined in (a) above, the corresponding kernels can be computed using any suitable kernel function that fulfills Mercer's conditions. In this work we consider the RBF kernel for the three components, with individual bandwidths determined to equalize the dynamic range of all kernels. Each component leads to a specific weighted adjacent matrix whose element $(i,j)$ represents the similarity between nodes $i$ and $j$ according to the corresponding measure, namely, spectral, spatial-spectral and spatial-2D. The spectral adjacent matrix is defined as~\cite{Camps-Valls-2007}
	\begin{equation}
		\begin{cases}
			W_{ij}^{\omega} = \exp(-\|\vx_i^{\omega}-\vx_j^{\omega}\|^2/2\sigma_{\omega}^2), &\text{for  } i \neq j\\
			W_{ii}^{\omega} = 0.
		\end{cases}
		\label{EqWf}
	\end{equation}
	\noindent where $\vx_i^{\omega}$ is the spectral representation of pixel $\vx_i$ (node of the graph).
	
	The spatial-spectral adjacent matrix is defined as~\cite{Camps-Valls-2007}
	\begin{equation}
		\begin{cases}
			W_{ij}^{s} = \exp(-\|\vx_i^{s}-\vx_j^{s}\|^2/2\sigma_{s}^2), &\text{for  } i \neq j\\
			W_{ii}^{s} = 0.
		\end{cases}
		\label{EqWd}
	\end{equation}
	\noindent where $\vx_i^{s}$ is a feature extracted from the spectra of the HI pixels in the area surrounding $\vx_i$.
	
	Finally, the proposed new spatial-2D adjacent matrix is defined as
	\begin{equation}
		\begin{cases}
			W_{ij}^{d} = \exp(-\|\vx_i^{d}-\vx_j^{d}\|^2/2\sigma_{d}^2), &\text{for  } i \neq j\\
			W_{ii}^{d} = 0.
		\end{cases} 
		\label{EqWd2}
	\end{equation}
	\noindent where $\vx_i^{d}$ is the 2-D vector with the coordinates of $\vx_i$ in the HI.
\end{itemize}

\subsection{The weighted-summation composite kernel}
A weighted-summation composite kernel considering the three pixel entities defined above can be constructed through a linear combination of the three weighted adjacent matrices defined in \eqref{EqWf}, \eqref{EqWd} and \eqref{EqWd2}, yielding a composite weight matrix $\mW$ as follows:
\begin{equation}
	\begin{split}
		W_{ij}(\vx_i, \vx_j) = &\beta W_{ij}^{\omega}(\vx_i^{\omega}, \vx_j^{\omega}) + \lambda W_{ij}^{d}(\vx_i^{d}, \vx_j^{d}) + \gamma W_{ij}^{s}(\vx_i^{s}, \vx_j^{s}).
	\end{split}
	\label{EqCompositeW}
\end{equation}

The values of $\sigma_{\omega}$, $\sigma_{d}$ and $\sigma_{s}$ are adjusted so that all elements of the three matrices $\mW^{\omega}$, $\mW^{s}$ and $\mW^{d}$ have the same value range. This can be accomplished by choosing these parameters such that
\begin{equation}
	\begin{split}
		&\frac{\max\left( \| \vx_i^{\omega} - \vx_j^{\omega}\|^2\right)}{2\sigma_{\omega}^2} = \frac{\max\left( \| \vx_i^{d} - \vx_j^{d}\|^2 \right)}{2\sigma_{d}^2} = \frac{\max\left( \| \vx_i^{s} - \vx_j^{s}\|^2 \right)}{2\sigma_{s}^2}.
	\end{split}
	\label{EqScaling}
\end{equation} 
Hence, the design of the proposed weighted-summation composite kernel according to \eqref{EqCompositeW} requires the choice of the three weights $\beta$, $\lambda$ and $\gamma$ (two weights for a convex sum), as well as one kernel bandwidth among $\sigma_{\omega}$, $\sigma_{d}$ and $\sigma_{s}$.

\section{Problem Solution} \label{sec_Solution}

The objective of a classification process using semi-supervised learning is to satisfy the so-called consistency assumption. As defined in~\cite{Zhou-2004}, consistency assumes that nearby points tend to belong to the same class, and that points on the same cluster or manifold belong to the same class. Hence, one needs an algorithm whose classification function is sufficiently smooth with respect to the inherent structure between a graph's labeled and unlabeled points.

To employ the algorithm proposed in~\cite{Zhou-2004} to the composite kernel approach, we assign the weight matrix $\mW$ defined in \eqref{EqCompositeW} to the required affinity matrix. 

Consider a classification problem with $m$ classes and an HI with $n$ pixels $\vx_i$ in the input space $\mathcal{X}$, $n_l$ of which are labeled with labels $y_i$, $i=1, \dots, n_l$, while the remaining $n_u$ pixels are unlabeled. For a graph-based approach, we define a graph $G = (V, E)$ on $\mathcal{X}$ where the vertex set $V$ is just $\mathcal{X}$ and the edges in set $E$ are weighted by $\mW$. The edge-connecting nodes $i$ and $j$ has an associated weight ${W_{ij}}$ which represent the weights of the edges in the data adjacency graph, connecting all labeled and unlabeled nodes. Let $\mathcal{F}$ be the set of $n \times m$ matrices with non-negative entries. By labeling each point $\vx_i$ as $y_i=\argmax_{j\leq m} F_{ij}$, a matrix $\mF=[\vf_1^T,\cdots,\vf_n^T]^T$ ${\in}$ $\mathcal{F}$  provides a classification that maps each input pixel $\vx_i$ in the dataset $\mathcal{X}$ to a vector $\vf_i \in \mathbb{R}^m$. Define also an $n\times m$  matrix $\mY \in \mathcal{F}$ with $Y_{ij}=1$ if $\vx_i$ is labeled as $y_i=j$ and $Y_{ij}=0$ otherwise. Notice that $\mY$ aligns with the initial labels assigned to the labeled pixels. 

The semi-supervised learning algorithm proposed in~\cite{Zhou-2004} with the affinity matrix \eqref{EqCompositeW} is then as follows:
\begin{itemize}
	\item[]\hspace{-4ex}Step 1: Define $\vx_i^{\omega}$, $i=1, \dots, n$ as the spectra of the $n$ HI pixels.
	\item[]\hspace{-4ex}Step 2: Determine the spatial-spectral pixel entities $\vx_i^s \in \mathbb{R}^{N_s}$, $i=1, \dots, n$, as the averages of the HI pixels in a window of size $w \times w$ around $\vx_i$.
	\item[]\hspace{-4ex}Step 3: Determine the spatial-2D pixel entities $\vx_i^d \in \mathbb{R}^2$, $i=1, \dots, n$, given by the coordinate vector of $\vx_i$ in the HI.
	\item[]\hspace{-4ex}Step 4: Choose a kernel bandwidth $\sigma_{d}$ to be used in \eqref{EqWd2} to determine $\mW^{d}$.
	\item[]\hspace{-4ex}Step 5: Determine kernel bandwidths $\sigma_s$ and $\sigma_{\omega}$ to satisfy \eqref{EqScaling}.
	\item[]\hspace{-4ex}Step 6: Construct the $n\times n$ matrices $\mW^s$ and $\mW^{\omega}$ in \eqref{EqWd} and \eqref{EqWd2}.
	\item[]\hspace{-4ex}Step 7: Choose the linear combination weights $\beta$, $\lambda$ and $\gamma$ to be used in \eqref{EqCompositeW}.
	\item[]\hspace{-4ex}Step 8: Form the $n\times n$ affinity matrix $\mW$ such that $W_{ij}(\vx_i, \vx_j) = \beta W_{ij}^{\omega}(\vx_i^{\omega}, \vx_j^{\omega}) + \lambda W_{ij}^{d}(\vx_i^{d}, \vx_j^{d}) + \gamma W_{ij}^{s}(\vx_i^{s}, \vx_j^{s})$.
	\item[]\hspace{-4ex}Step 9: Construct the matrix $\mS = \mD^{-1/2} \mW \mD^{-1/2}$ in which $\mD$ is a diagonal matrix with elements $D_{ii} = \sum_{j} W_{ij}$.
	\item[]\hspace{-4ex}Step 10: Construct the $n\times m$ matrix $\mY$ with $Y_{ij}=1$ if $\vx_i$ is labeled as $y_i=j$ and $Y_{ij}=0$ otherwise.
	\item[]\hspace{-4ex}Step 11: Iterate $\mF(k+1) = \alpha \mS\mF(k) +  (1 - \alpha)\mY$ until convergence, where $\alpha$ is a parameter in $(0,1)$.\\
\end{itemize}

It is straightforward to show that $\mF(k)$ in Step 11 converges to
\begin{equation}
	\mF^{*}   = \lim_{k\rightarrow\infty } \mF(k)= (1-\alpha) (\mI-\alpha \mS)^{-1}\mY
	\label{EQF} 
\end{equation}
\noindent for $\alpha$ in $(0,1)$.

A pseudocode for the proposed classification method is presented in Algorithm~\ref{alg:method}.

\begin{algorithm}[!htb] 
	\SetAlgoLined
	\KwIn{hyperspectral image pixels $\vx_i^{\omega}$, $i=1, \dots, n$, parameters $\beta$, $\lambda$, $\gamma$, $\sigma_{d}$, $\alpha$.}
	Determine $\vx_i^s \in \mathbb{R}^{N_s}$, $i=1, \dots, n$.\\
	Determine $\vx_i^d \in \mathbb{R}^2$, $i=1, \dots, n$\\
	$\sigma_{s}^2 = \sigma_{d}^2 \frac{\max\left( \| \vx_i^{s} - \vx_j^{s}\|^2 \right)}{\max\left( \| \vx_i^{d} - \vx_j^{d}\|^2\right)}$\\
	$\sigma_{\omega}^2 = \sigma_{d}^2 \frac{\max\left( \| \vx_i^{\omega} - \vx_j^{\omega}\|^2 \right)}{\max\left( \| \vx_i^{d} - \vx_j^{d}\|^2\right)}$\\
	$\mW^{\omega} \leftarrow$ Eq. \eqref{EqWf}\\
	$\mW^s \leftarrow$ Eq. \eqref{EqWd}\\
	$\mW^d \leftarrow $ Eq. \eqref{EqWd2}\\
	$\mW = \beta \mW^{\omega} + \lambda \mW^d + \gamma \mW^s$ \\
	$\mS = \mD^{-1/2} \mW \mD^{-1/2}$, $D_{ii} = \sum_{j} W_{ij}$ \\
	$\mY \leftarrow $ $Y_{ij}=1$ if $y_i=j$ and $Y_{ij}=0$ otherwise.\\
	$\mF^{*} = (1-\alpha) (\mI-\alpha \mS)^{-1}\mY$\\
	\For{$i=n_l+1,\dots,n$}{
		$y_i=\argmax_{j\leq m} F_{ij}^*$.
	}
	\Return{\rm{Classes} $y_i$ \rm{for each} $\vx_i, i > n_l$};
	\caption{\textit{Semi-supervised HI classification}} \label{alg:method}
\end{algorithm}

\section{Experimental Results and Discussions} \label{sec_Simulations}

To evaluate the performance of the proposed composite kernel approach, we applied Algorithm~\ref{alg:method} to semi-supervised classification of HIs. Simulation experiments have been realized using two well known real HIs, namely, the AVIRIS Indian Pines and the ROSIS Pavia University datasets to allow for comparison with results previously published without available software.

\begin{figure}[!htb]
	\centering
	\includegraphics[width=0.7\textwidth]{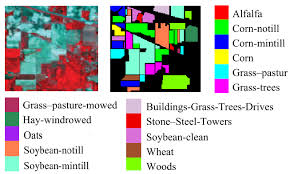}
	\label{subimagem}
	\caption{Indian Pines scene displayed in false color using the red, green, and blue bands, together with the ground reference map and class legend.}
	\label{IP}
\end{figure}

The AVIRIS dataset, comprising 224 spectral bands with spatial dimensions of $145 \times 145$ pixels, was acquired over the Indian Pines site in 1992. The data have a spatial resolution of $20\,$m and a spectral resolution of $10$\,nm, spanning the $400$–$2500$\,nm wavelength range. After discarding noisy and water absorption bands, 200 effective bands were retained. Figure \ref{IP} depicts the scene, which is predominantly composed of agricultural areas encompassing 16 land-cover classes. Notably, corn and soybean fields display differences related to soil tillage practices.

\begin{figure}[!htb]
	\centering
	\includegraphics[width=0.7\textwidth]{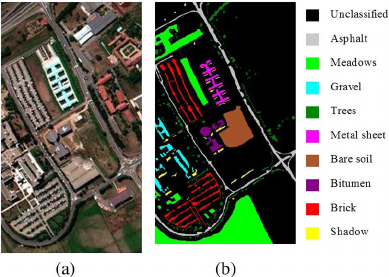}
	\caption{Pavia University (PU) image. (a) False-color composition of three bands (b) Ground reference information and the corresponding class legend.}
	\label{PU}
\end{figure}

The ROSIS (Reflective Optics System Imaging Spectrometer) dataset, with high spatial resolution and comprising 115 spectral bands, was acquired over Pavia University (PU) in Italy in 2002. The scene has spatial dimensions of $610 \times 340$ pixels, a spatial resolution of $1.3$~m, and spectral coverage ranging from $430$~nm to $860$~nm. After removing noisy bands, 103 bands were retained for the experiments. Figure \ref{PU} illustrates the scene, which contains diverse urban and vegetation classes, providing a challenging benchmark for hyperspectral classification algorithms. The images illustrate a false-color composition generated from three bands, along with ground reference information and the associated class legend.

The following experiments illustrate the benefits of integrating the spectral graph-based method with an auxiliary graph that incorporates the two-dimensional spatial distances between pixels in the image. The different experiments have been tailored so that the results can be directly compared to results previously reported in the literature.

\subsection{First Experiment}

In this experiment, we employed a subset of Indian Pines comprising pixels [27–94] $\times$ [31–116], resulting in a $68 \times 86$ scene containing four labeled classes (background pixels were excluded from classification): “corn-no till” (1008), “Grass/Trees” (732), “Soybeans-no till” (727), and “Soybeans-min” (1926). This scenario is particularly challenging for land-cover classification, as the main crops in the  -- primarily corn and soybeans -- were at early growth stages, especially given the moderate spatial resolution of $20\,$m. The subset is used to evaluate the performance of the proposed kernel composition method in comparison with the results presented in \cite{Camps-Valls-2007}. The parameters of the algorithm were adjusted based on a grid search with $\alpha$ values in [0.1, 0.9], $\sigma_d^2$ values in the the interval [10, 200], $\beta$ and $\lambda$ in [0.1, 0.9] and $\gamma = 1-\lambda - \beta$. For each combination of $\alpha$ and $\sigma_d^2$ values, parameters  $\sigma_s^2=7.57$ and $\sigma_{\omega}^2=8.36$ were determined from \eqref{EqScaling}. The following parameters were used in the simulations: $\alpha=0.5$,  $\sigma_d^2=110$, $\sigma_s^2=7.57$ and $\sigma_{\omega}^2=8.36$. The reported accuracy results were obtained by averaging over 50 independent runs. The spatio-spectral samples ${\vx}_i^{s}$ were computed as the average over a $3 \times 3$ window surrounding each pixel ${\vx}_i$ for each band, thereby defining $W_{ij}^s$ as specified in equation \eqref{EqWd}.

Table \ref{subIP} presents the accuracy results for different amounts of initially known labels and combinations of spectral, spatial-spectral and 2D-spatial kernels. The challenge of balancing the weights $\beta$, $\lambda$ and $\gamma$ can lead to one type of information being favored over the others. It is worth noting that the results reported in \cite{Camps-Valls-2007}, also included in Table \ref{subIP}, correspond to the best-performing kernel compositions among the various configurations tested by the authors in \cite{Camps-Valls-2007}. Parameter variations in Equation (\ref{EqCompositeW}) were examined to assess the effectiveness of the method, as well as the corresponding value when one of these weights was omitted in the context of composite kernel-based image classification. 

To derive a consistent framework for analyzing possible relationships, a careful adjustment of the parameters allows us to better understand how changes in $\lambda$, $\beta$, and $\gamma$ influence the method's accuracy when using the spatial-spectral kernel or not. To this end we started with a convex combination such that $\lambda+\beta+\gamma=1$. We recall from \eqref{EqCompositeW} that $\lambda$ is the weight associated with the contribution of the newly proposed spatial-2D kernel to the weight matrix $\mW$. Then, to compare the results obtained using the 3-component convex combination with a 2-component combination including only one spectral and one spatial-2D term we eliminate one of the spectral terms while keeping the ratio of the weights of the remaining terms unaltered. For instance, to remove the term $\mW^s$ we determine the ratio $k = \lambda/\beta$ of the original combination. Then, the weights $\lambda_e$ and $\beta_e$ of the ``equivalent'' 2-component combination are determined so that  $\lambda_e + \beta_e=1$ and $\lambda_e/\beta_e=k$. This result is obtained making $\lambda_e=k/(k+1)$ and $\beta_e=1/(k+1)$. Likewise, when eliminating the term $\mW^{\omega}$ from the original combination, we define $k = \lambda/\gamma$, leading to $\lambda_e=k/(k+1)$ and $\gamma_e=1/(k+1)$.

Table~\ref{subIP} shows different results obtained using the described methodology. Different combinations of coefficient values $\beta$, $\lambda$, and $\gamma$, as well different amounts of initially labeled pixels were considered. The labeled pixels initially selected for each class were chosen randomly for each simulation run. In Case 1, $\mW^d$ had the same weight as the two spectral terms combined. The obtained classification accuracies for all four classes were better than the best results reported in \cite{Camps-Valls-2007} for all kernel composition used. In Case 2, the spectral-spatial component $\mW^s$ was removed while maintaining the relative weights of $\mW^{\omega}$ and $\mW^d$ as explained above. An accuracy improvement was verified, indicating that the proposed spatial-2D kernel provided better spatial information than the spectral-spatial kernel. In Case 3 we removed the contribution of the pure spectral component $\mW^{\omega}$, while again keeping the ratio of the other two contributions unaltered. Comparison of the accuracies obtained for these three cases and the results reported in \cite{Camps-Valls-2007} indicates that the proposed spatial-2D component provided important spatial information. They also indicate that the simple spectral component $\mW^{\omega}$ should be preferred over the spatial-spectral component $\mW^s$ if only one of them is to be employed.

Differently from Case~1, Case~4 considered a situation in which the contribution of the spatial-2D component $\mW^d$ was much smaller than the combined contributions of the two spectral-based components $\mW^{\omega}$ and $\mW^s$. The accuracy results obtained were clearly inferior to those obtained in Case~1, Case~2, Case~3 and those reported in \cite{Camps-Valls-2007}. Case~5, derived form Case~4 by eliminating the spectral-spatial component $\mW^s$ while maintaining the ratio $\lambda/\beta$, led to a significant improvement in classification accuracy, supporting the previous indication that the use of $\mW^{\omega}$ and $\mW^d$ provided a better combination of spectral and spatial information for classification purposes than the combination of $\mW^{\omega}$ with $\mW^s$. Comparison of all cases  and the results from \cite{Camps-Valls-2007} clearly indicate the relevance of the proposed spatial-2D to the classification accuracy, with the parameters in Case~2 leading to the overall best results for all classes. 

Finally, a comparison of the results in Case~2 and Case~5 showed a small sensitivity of the accuracy results to a significant variation of the relative weights applied to $\mW^{\omega}$ and $\mW^d$, as well as the importance of employing the proposed spatial-2D component.

\begin{table}[!htb]
	\caption{ Classification accuracy ($\%$) for different numbers of labeled samples per class - Indian Pines Image}
	\renewcommand{\arraystretch}{1.1}
	\begin{center}
		\begin{tabular}{|>{\arraybackslash}p{1.6cm}|
				>{\centering\arraybackslash}p{0.9cm}|
				>{\centering\arraybackslash}p{0.9cm}|
				>{\centering\arraybackslash}p{0.9cm}|
				>{\centering\arraybackslash}p{0.9cm}|
				>{\centering\arraybackslash}p{0.9cm}|}
			\hline
			Number of\newline labeled points per class &\multirow{3}{*}{$3$} &\multirow{3}{*}{$5$} &\multirow{3}{*}{$15$} &\multirow{3}{*}{$30$}\\ \cline{1-1}
			\hline
			\hline
			Composite\newline Kernel &\multicolumn{4}{|c|}{\multirow{2}{*}{Accuracy (\%)}} \\ \cline{1-1} 
			\hline
			${\bf Case~1}$ &  &  &  &   \\
			$\lambda=0.50$ &  &  &  &   \\
			$\beta=0.25$ & $71.10 $ & $75.88$ & $82.33 $ & $ 85.40$ \\ 
			$\gamma=0.25$ &  &  &  &   \\
			\hline
			${\bf Case~2 }$ & & & &  \\
			$\lambda_e=0.67$ & $ {\bf 73.46}$ & ${\bf 79.14}$ & $ {\bf 85.27}$ & $ {\bf 87.00}$ \\
			$\beta_e=0.33$  & & & &  \\
			\hline
			${\bf  Case~3 }$ & & &  & \\
			$\lambda_e=0.67$  & $69.20$ & $77.34 $ &  $83.33$ &  $84.68$ \\ 
			$\gamma_e=0.33$ & & &  &\\
			\hline
			${\bf Case~4 }$ & & &  & \\
			$\lambda=0.10$ & & &  & \\ 
			$\beta=0.45$  & $ 52.60 $ & $57.91 $ & $65.34 $ & $ 66.71$ \\
			$\gamma=0.45$ & & &  &\\
			\hline
			${\bf  Case~5 }$ &  & & & \\
			$\lambda_e=0.18$   & $71.54 $ & $77.42$ & $ 84.83 $ & $ 85.80$ \\
			$\beta_e=0.82$ & &  & & \\
			\hline
			{\bf Ref. \cite{Camps-Valls-2007} }  & $66.73$ & $67.13$ & $79.49$ & $84.99$ \\
			\hline
		\end{tabular}
	\end{center}
	\label{subIP}
\end{table}

The results in Table~\ref{subIP} show that, regardless of how small the value of $\lambda$ associated with the 2D-spatial kernel was chosen, there was no strong dependency on the combination of the spectral kernels $\mW^{\omega}$ and $\mW^s$ for the Indian Pines dataset. This outcome suggests a reduction in computational complexity, as the contribution of the spatial–spectral kernel $\mW^s$ can be reasonably disregarded. Specifically for the Indian Pines image, this behavior can be attributed to the averaging process used in the construction of the matrix $\mW^s$ from the original image cube, which does not substantially modify the mean spectral characteristics evaluated in a window about most of the pixels in the image due to the occurrence of large areas with pixels of the same class.

\subsection{Second Experiment}

We used the Pavia University (PU) image for this experiment. To ensure a balanced assessment, five classes (Gravel (2099), Trees (3064), Painted metal sheets (1345), Bitumen (1330) and Shadows (947)) with a moderate amount of labeled data were selected for the evaluation of the method. It should be emphasized that simulations conducted with alternative class combinations did not result in statistically significant differences in accuracy. As in the first example, the spatio-spectral samples ${\vx}_i^{s}$ were computed as the average over a $3 \times 3$ window surrounding each pixel ${\vx}_i$ for each band. The reported accuracies are the result of an average over 50 independent runs. Also, the algorithm parameters were adjusted based on a grid search with the same parameter intervals employed in Experiment~1. For each combination of $\alpha$ and $\sigma_d^2$ values, parameters  $\sigma_s^2$ and $\sigma_{\omega}^2$ were determined from \eqref{EqScaling}. The parameters used in the simulations were $\alpha=0.5$,  $\sigma_d^2=10$, $\sigma_s^2=0.51$ and $\sigma_{\omega}^2=0.63$. 

Table \ref{subPU} presents accuracy results obtained for different choices of the linear combination coefficients $\beta$, $\lambda$ and $\gamma$, and for different amounts of initially labeled pixels. The pixels initially labeled for each class were chosen randomly for each simulation run. 

We proceeded as in the first example in that the parameters of the 2-coefficient linear combinations were obtained from a previously tested 3-coefficient combination by maintaining the ratio of the coefficients of the preserved terms and keeping the convex nature of the sum.   

\begin{table}[!htb]
	\caption{ Classification accuracy (\%) to different numbers of labeled samples per class - PU Image}
	\renewcommand{\arraystretch}{1.0}
	\begin{center}
		\begin{tabular}{|>{\arraybackslash}p{1.6cm}|
				>{\centering\arraybackslash}p{0.9cm}|
				>{\centering\arraybackslash}p{0.9cm}|
				>{\centering\arraybackslash}p{0.9cm}|
				>{\centering\arraybackslash}p{0.9cm}|
				>{\centering\arraybackslash}p{0.8cm}|}
			\hline
			Number of\newline labeled points per class &\multirow{3}{*}{$3$} &\multirow{3}{*}{$5$} &\multirow{3}{*}{$15$} &\multirow{3}{*}{$30$}\\ \cline{1-1}
		\hline
		\hline
		Composite\newline Kernel &\multicolumn{4}{|c|}{\multirow{2}{*}{Accuracy (\%)}} \\ \cline{1-1} 
		\hline
		${\bf Case 1}$ &  &  &  &   \\
		$\lambda=0.50$ &  &  &  &   \\
		$\beta=0.25$ & $97,16 $ & $97.75$ & $97.93 $ & $ 98.03$ \\ 
		$\gamma=0.25$ &  &  &  &   \\
		\hline
		${\bf Case 2 }$ & & & &  \\
		$\lambda_e=0.67$ & $ 90.50$ & $92.13$ & $ 94.26$ & $ 94.79$ \\
		$\beta_e=0.33$  & & & &  \\
		\hline
		${\bf  Case 3 }$ & & &  & \\
		$\lambda_e=0.67$  & $ 98.13$ & $ {\bf 98.47} $ &  ${\bf 98.64}$ &  ${\bf 98.72}$ \\ 
		$\gamma_e=0.33$ & & &  &\\
		\hline
		${\bf Case 4 }$ & & &  & \\
		$\lambda=0.10$ & & &  & \\ 
		$\beta=0.45$  & $ 97.26 $ & $97.44$ & $97.74 $ & $ 97.90$ \\
		$\gamma=0.45$ & & &  &\\
		\hline
		${\bf  Case 5 }$ &  & & & \\
		$\lambda_e=0.18$   & $89.60 $ & $92.26$ & $ 92.21 $ & $ 92.71 $\\
		$\beta_e=0.82$ & &  & & \\
		\hline
		${\bf Case 6}$ & &  & & \\
		$\lambda=0.25$  &  & & & \\
		$\beta=0.50$   & $95.77 $ & $96.14$ & $ 97.06 $ & $97.38 $ \\
		$\gamma=0.25$ & &  & &\\
		\hline
		${\bf Case 7}$ & &  & & \\
		$\lambda=0.50$  &  & & & \\
		$\beta=0.0$   & $ {\bf 98.18}$ & $98.30$ & $ 98.56 $ & $ 98.59$ \\
		$\gamma=0.50$ & &  & &\\
		\hline
		${\bf Case 8}$ & &  & & \\
		$\lambda=0.50$  &  & & & \\
		$\beta=0.50$   & $90.50 $ & $92.30 $ & $ 93.23 $ & $ 93.27$ \\
		$\gamma=0.0$ &  &  & &\\
		\hline
	\end{tabular}
\end{center}
\label{subPU}
\end{table}

An analysis of Table~\ref{subPU} reveals, unlike the results obtained for the Indian Pines image, a reduced sensitivity of the classification performance to the choice of weights chosen for $\mW^d$, $\mW^s$ and $\mW^{\omega}$. Overall, the results exhibited only small variations across the evaluated cases. Notably, Case~3, in which the composite kernel did not include a contribution of $\mW^{\omega}$, has yielded the best classification accuracies for most of the tested cases. This shows that the spectral-spatial pixel representation provided important spectral information. It is noteworthy that this trend is also reflected in the class-wise accuracy, as can be verified in Table~\ref{classesPU}, which presents accuracy results obtained per class for selected combinations of the parameters $\lambda$, $\beta$, and $\gamma$. Notice that a comparison of cases 2 and 3 in Table~\ref{classesPU} indicates that for this image the spatial-spectral contributions $\vx^s$ provided a spectral information that is more relevant than the information provided by the pixels $\vx^{\omega}$. Also, comparison of the results obtained for cases 1 and 4 indicates that employing the newly proposed pixel entity $\vx^d$ can lead to an improvement in classification accuracy even for this image.

\begin{table}[!htb]
\caption{ Classification accuracy for the different classes - PU Image.}
\renewcommand{\arraystretch}{1.0}
\begin{center}
	\begin{tabular}{|>{\arraybackslash}p{3.5cm}|
			>{\arraybackslash}p{4.5cm}|
			>{\centering\arraybackslash}p{3.5cm}|
			>{\centering\arraybackslash}p{3.5cm}|
			>{\centering\arraybackslash}p{3.5cm}|
			>{\centering\arraybackslash}p{3.5cm}|}
		\hline
		\multicolumn{2}{|c|}{Accuracies (\%) for 5 labeled points per class} \\
		\hline
		${\bf Case~1}$ &    Gravel = 97.48 \\
		$\lambda=0.50$ &  Trees= 99.74  \\ 
		$\beta=0.25$ & Painted metal sheeets= 99.48\\
		$\gamma=0.25$   & Bitumen = 94.64 \\ &  Shadows = 97.38\\
		\hline
		${\bf Case~2}$ &  Gravel = 83.24   \\
		$\lambda=0.50$ &  Trees= 99.74  \\ 
		$\beta=0.50$ &  Painted metal sheeets= 99.44 \\
		$\gamma=0.0$   & Bitumen = 86.03 \\ & Shadows = 87.57 \\
		\hline
		${\bf Case~3}$ &   Gravel = 97.11  \\
		$\lambda=0.50$ &  Trees= 99.62  \\ 
		$\beta=0.0$ & Painted metal sheeets= 99.96  \\
		$\gamma=0.50$   & Bitumen = 94.63 \\ & Shadows = 99.66 \\
		\hline
		${\bf Case~4}$ &  Gravel = 95.80   \\
		$\lambda=0.0$ &  Trees=  99.74  \\ 
		$\beta=0.50$ & Painted metal sheeets= 99.44 \\
		$\gamma=0.50$ & Bitumen = 93.72 \\ & Shadows = 96.50 \\
		\hline
	\end{tabular}
\end{center}
\label{classesPU}
\end{table}

\subsection{Third Experiment}

This experiment compared the performances of the proposed method and the method of \cite{Shao-discriminant-2016}. To this end we considered Class 3 (Corn-Min) and Class 12 (Soybeans-clean) of the Indian Pines image. Moreover, since only a subset of the pixels belonging to each class in the original image was considered in the experiment described in~\cite{Shao-discriminant-2016}, we employed the same number of pixels for each of the two classes to enable a fair comparison. As a result, 270 pixels were selected at random from Class 3, out of a total of 830 pixels, and 261 pixels were chosen (also randomly) from Class 12, which contained 593 pixels.

We performed a grid search as in the previous experiments to determine the parameters of the proposed algorithm. After several trials, we verified that excellent classification results were obtained applying only the spectral kernel $W_{ij}^{\omega}$ and the 2D-spatial kernel $W_{ij}^d$. Moreover, this choice reduced the computational complexity required when compared to a solution including the spectral–spatial kernel $W_{ij}^s$. The overall (OA) results shown in Table \ref{experimento3} for different amounts of initially labeled samples were obtained using the following set of parameters: $\alpha=0.5$, $\sigma_d^2=50$, $\sigma_{\omega}^2=1.34$, $\lambda=0.55$ and $\beta = 1 - \lambda$.

We observe that our approach provided more accurate results than those reported in \cite{Shao-discriminant-2016} in all cases.

\begin{table}[!htb]
\caption{Classification accuracy for the Indian-Pines Image. Results for the proposed method and for \cite{Shao-discriminant-2016}. }
\renewcommand{\arraystretch}{1.5}
\begin{center}
\begin{tabular}{|>{\centering\arraybackslash}p{2.0cm}|
		>{\centering\arraybackslash}p{2.5cm}|
		>{\centering\arraybackslash}p{2.5cm}|}
	\hline
	\multicolumn{3}{|c|}{Classes 3 e 12 - OA($\%$)}  \\
	\hline
	Number of labeled points& Reference \cite{Shao-discriminant-2016} &  Proposed Method \\
	\hline
	3   &  $95.50$ & {\bf 97.23}\\
	\hline
	5   &  $97.00$ & {\bf 98.41}\\
	\hline
	10  & $98.00$ & {\bf 98.74} \\
	\hline
	15  & $98.20$ & {\bf 99.13}\\
	\hline
\end{tabular}
\end{center}
\label{experimento3}
\end{table}

Table~\ref{experimento3A}  reports the accuracy results obtained for each individual class. Excellent results were observed for both classes regardless of the number of samples initially labeled per class. Class-wise accuracy results are not presented in \cite{Shao-discriminant-2016}.

\begin{table}[!htb]
\caption{ Third Experiment. Results of acccuracy per class.} 
\renewcommand{\arraystretch}{1.5}
\begin{center}
\begin{tabular}{|>{\centering\arraybackslash}p{1.0cm}|
		>{\centering\arraybackslash}p{1.3cm}|
		>{\centering\arraybackslash}p{1.3cm}|
		>{\centering\arraybackslash}p{1.3cm}|
		>{\centering\arraybackslash}p{1.3cm}|}
	\hline
	\multicolumn{5}{|c|}{Number of labeled points} \\
	\hline
	Classes & $3$ & $5$ & $10$ & $15$ \\
	\hline
	{\bf 3}   &  $98.4$ & $99.79$ & $100$ & $100$\\
	\hline
	{\bf 12}   & $96.06$ & $97.03$ & $97.45$ & $98.23$ \\
	\hline
\end{tabular}
\end{center}
\label{experimento3A}
\end{table}

\subsection{Fourth Experiment}

In this experiment, we compared our proposed method with that of  \cite{Shao-discriminant-2016} when all 16 classes present in the Indian Pines image were considered. In \cite{Shao-discriminant-2016}, the amounts of pixels used to represent each class were as follows: class 1 = $46$, class 2 = $100$, class 3 = $270$, class 4 = $234$, class 5 = $63$, class 6 = $101$, class 7 = $28$, class 8 = $478$, class 9 = $20$,  class 10 = $66$,  class 11 = $122$,  class 12 = $261$,  class 13 = $205$,  class 14 = $117$,  class 15 = $291$ and class 16 = $93$. For a fair comparison, we used the same scenario for the proposed method.  After a grid search as described in the previous experiments, the following parameters were used in the simulations: $\alpha=0.5$, $\lambda=0.55$,  $\beta=0.45$, $\gamma=0$, $\sigma_d^2=20 $ and $\sigma_{\omega}^2=0.74$.

Table \ref{table3} presents comparative results for accuracy by class and overall accuracy (OA) obtained using the proposed method and the results reported in \cite{Shao-discriminant-2016}. The experiment considered $15$ available labeled samples per class, and the reported results correspond to the average of $20$ runs.  

We can observe that the proposed method was superior in classification accuracy for all classes but for class $13$ when compared to the method in \cite{Shao-discriminant-2016}. The overall average was significantly superior.  

\begin{table}[!htb]  
\caption{Comparison between the proposed method and \cite{Shao-discriminant-2016} for the Indian-Pines Image.}
\renewcommand{\arraystretch}{2}
\begin{center}
\begin{tabular}{|>{\centering\arraybackslash}p{2.0cm}|
		>{\centering\arraybackslash}p{2.5cm}|
		>{\centering\arraybackslash}p{2.5cm}|}
	\hline
	Classes & \cite{Shao-discriminant-2016}- DSR -Graph  &  Proposed Method \\
	\hline
	1 &$87.3$ & {\bf 100}\\
	\hline
	2& $84.52$ & {\bf 100}\\
	\hline
	3 & $88.5$ & {\bf 91.14} \\
	\hline
	4 & $76.62$ & {\bf 96.18}\\
	\hline
	5 &  90.93 &  {\bf 100}\\
	\hline
	6 & $99.53$ & {\bf 100}\\
	\hline
	7 & $91.36$ & {\bf 100}\\
	\hline
	8 & $73.16$ & {\bf 99.88}\\
	\hline
	9 & ${\bf 100}$ & ${\bf  100}$\\
	\hline
	10 & $69.11$ & {\bf 95.73}\\
	\hline
	11 & $61.44$ & {\bf 100}\\
	\hline
	12 & $65.60$ & {\bf 90.10}\\
	\hline
	13 & {\bf 99.11} & 92.65\\
	\hline
	14 & $87.79$ & {\bf 100}\\
	\hline
	15 & $43.67$ & {\bf 99.44}\\
	\hline
	16 & $99.31$ & {\bf 100}\\
	\hline
	OA & $79.63$ & {\bf 96.74} (0.77)\\
	\hline
\end{tabular}
\end{center}
\label{table3}
\end{table}

\begin{figure}[!htb]
\centering
\includegraphics[width=0.7\textwidth]{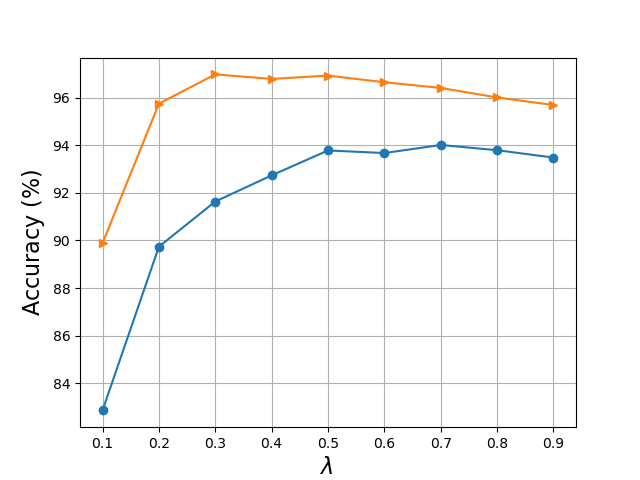}
\caption{Fourth experiment. Overall accuracy (OA)  for different values of $\lambda$. Indian-Pines image. Blue curve: 5 labeled samples per class. Orange curve: 15 labeled samples per class.}
\label{graf}
\end{figure}

Figure \ref{graf} shows the overall accuracy results obtained with the proposed method for different values of $\lambda$ in equation (\ref{EqCompositeW}), with $\beta = 1 - \lambda$, and $\gamma=0$. The graph shows curves related to the use of $5$ (blue curve) and $15$ (orange curve) initially labeled samples per class. These results indicate a strong robustness to variations in $\lambda$ in both scenarios.

\subsection{Fifth Experiment}

In this experiment, we compared the performance of the proposed method with the method presented in \cite{Ma-graph-based-2016}. To this end we used the same scenario described in \cite{Ma-graph-based-2016}, choosing the same nine classes from the Indian-Pines image, namely, classes $2$, $3$, $5$, $6$, $8$, $10$, $11$, $12$ and $14$. As in \cite{Ma-graph-based-2016}, we randomly selected $40\%$ of the $9234$ pixels available in the nine classes, that is, a total of $3738$ pixels. After a grid search as done for the previous experiments, the parameters used in the simulations were $\lambda=0.45$, $\beta = 1- \lambda$, $\gamma = 0$, $\alpha=0.5$, $\sigma_d^2=20 $, and $\sigma_{\omega}^2=0.924$. Twenty labeled samples per class were used and we calculated the average of 20 runs to obtain the accuracy results.

Table \ref{table4} compares the results obtained using the proposed method with the best results reported in \cite{Ma-graph-based-2016}. Except for a small difference in the result for Class 3, the proposed method has led to superior accuracies for all other classes. 

\subsection{Discussion}

An interesting aspect of the results obtained in the experiments is the relative importance of the two spectrally based similarity measures employed in the spectral and in the spectral-spatial components of the composite kernel. The experiments have evidenced that the relative effectiveness of each similarity measure depends on characteristics of the HI being classified. 

\begin{figure}[!htb]
\centering
\includegraphics[width=0.7\textwidth]{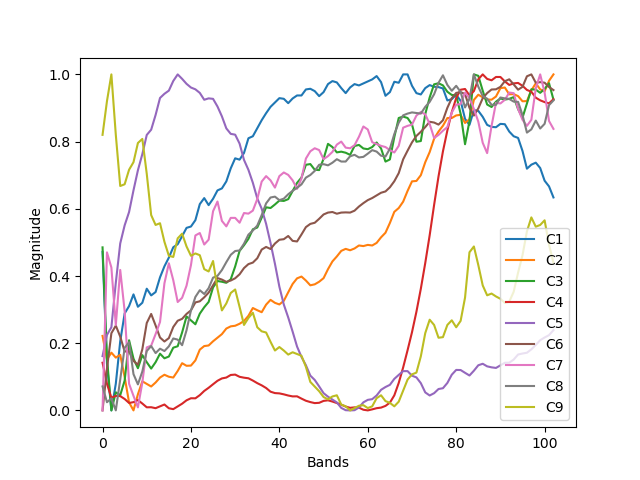}
\caption{Mean value of the spectra of each class of the Pavia HI.}
\label{Media_pavia}
\end{figure}

Fig.~\ref{Media_pavia} shows the spectral curves corresponding to the 103 bands of the pixel classes in the Pavia hyperspectral image. An inspection of  these curves indicates that, for most wavelengths, the spectral magnitudes exhibit substantial separation. This property supports the simulation results, wherein the exploitation of local neighborhoods through spatial windowing has the potential to yield a meaningful improvement in classification accuracy when combined with the adopted 2D spatial distance.

\begin{figure}[!htb]
\centering
\includegraphics[width=0.7\textwidth]{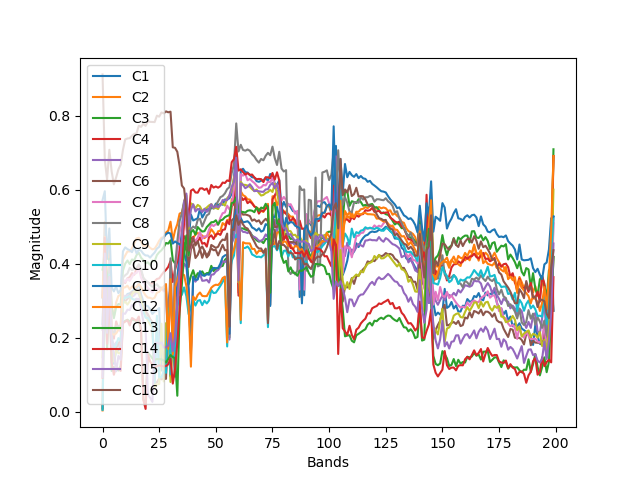}
\caption{Mean value of the spectra of each class of the Indian Pines HI.}
\label{Media_Indian}
\end{figure}

Conversely, this behavior is not observed in the Indian Pines dataset. As illustrated in Fig.~\ref{Media_Indian}, the spectral signatures across its 200 bands remain tightly clustered, with minimal inter-class separation. This observation is consistent with the simulation outcomes, which show that window-based nearest-neighbor exploration provides limited gains in accuracy for the kernel formulation employed in this study.

\begin{table}[!htb]
\caption{ Comparison between the proposed method and the reference \cite{Ma-graph-based-2016} for the Indian-Pines Image.}
\renewcommand{\arraystretch}{2}
\begin{center}
\begin{tabular}{|>{\centering\arraybackslash}p{2.0cm}|
		>{\centering\arraybackslash}p{2.5cm}|
		>{\centering\arraybackslash}p{2.5cm}|}
	\hline
	Classes& Reference \cite{Ma-graph-based-2016}  & Proposed Method  \\
	\hline
	2   & 93.16 & {\bf 98.92} \\
	\hline
	3   & {\bf 98.48} & 97.16\\
	\hline
	5  & 95.38 & {\bf 97.03} \\
	\hline
	6  &  {\bf 100} & {\bf 100} \\
	\hline
	8   & {\bf 100} & {\bf 100}\\
	\hline
	10   &93.99  & {\bf 97.12} \\
	\hline
	11  & 91.76 & {\bf 99.81} \\
	\hline
	12  & 92.56 & {\bf 100} \\
	\hline
	14   & {\bf 100} & {\bf 100} \\
	\hline
	{\bf OA}$\%$ & 95.28 & {\bf 98.92} (0.64)   \\
	\hline
\end{tabular}
\end{center}
\label{table4}
\end{table}

\section{Conclusions} \label{sec_Conclusions}

This work has proposed an alternative solution leveraging spatial information in graph-based semi-supervised (SS) transductive hyperspectral image classification. The consideration of pixel spatial properties in SS classification of hyperspectral images has been a subject of many studies with the objective of performance improvement. However, most existing solutions obtain the spatial properties as a function of the spectra of pixels in a spatial region about each HI pixel. While this approach adds some spatial information to the learning process, such information is still spectral in nature. One motivation for using such spectral-based spatial information was the use of cross-information kernels which required that spectral and spatial representations had the same dimension. Inspired by works in genomic data fusion which employed a kernel learning framework combining different types of data descriptions, as well as by works on conditional random fields (CRF) defined over pixels or image regions to refine pixel labeling, we proposed the addition of a new term to the typical composite kernel which considers exclusively spatial properties of the pixels. Our approach is simple to apply, as it does not require spectral operations about each pixel of the image, and can be directly applied to one of the simplest and most effective graph-based learning approaches. Experimental results using real images showed that the proposed composite kernel adds significant information to the learning process, yielding better classification results when compared to results previously reported in the literature.  The new results also help in understanding the relative contributions of the two spectral-based contributions to the composite kernel, which depend on the properties of the HI being classified. It is our conjecture that the proposed approach can also be applied to generate new directed-graph-based classification methods which emphasize the influence of the labeled samples in classifying the unlabeled ones. This is a possible direction for future work.

\bibliographystyle{elsarticle-num}
\bibliography{bibliografia}

\end{document}